\documentstyle[aps,epsf,floats,titlepage]{revtex}
\input epsf
\begin{document}
\title{Spin-dynamic field coupling in strongly THz driven semiconductors :
        local inversion symmetry breaking}
\author
{Kristinn Johnsen}
\address{
NORDITA, Blegdamsvej 17, DK-2100 Copenhagen \O, Denmark}
\date{\today}
\maketitle
\begin{abstract}
We study theoretically the optics in undoped direct gap semiconductors which are strongly
driven in the THz regime. We calculate the optical sideband generation due to nonlinear mixing of
the THz field and the near infrared probe. Starting with an inversion symmetric microscopic
Hamiltonian we include the THz field nonperturbatively using non-equilibrium Green function
techniques. We find that a self induced relativistic spin-THz field coupling
 locally breaks the inversion symmetry, resulting in the formation of odd sidebands which 
otherwise are absent.
\end{abstract}
%
\begin{widetext}
\section{INTRODUCTION}
Optical properties of semiconductors are sensitive to external conditions
such as strong static electric fields being applied to the sample. This
was predicted more than 40 years ago by Franz and Keldysh \cite{Franz-Keldysh}
and leads to the Franz-Keldysh effect. The effect manifests itself in finite
absorption within the gap, near the band edge, and modulation of the above
gap absorption spectrum.
The effect can be understood in terms of tunneling assisted absorption.
This work was generalized to oscillating
external fields, $\vec E(t) = \vec E\cos (\Omega t)$,
by Yacoby \cite{Yacoby} 30 years ago. Yacoby concluded that
similar effects to the Franz-Keldysh effect manifest provided that the
ponderomotive energy,
\begin{equation}
E_f = \hbar\omega_f = \frac{e^2E^2}{4m_r\Omega^2},
\label{pounderomotive}
\end{equation}
magnitude is similar to the photon energy, $\hbar\Omega$, of the external field, i.e.\
\begin{equation}
\gamma = \frac{\omega_f}{\Omega} = \frac{e^2E^2}{4\hbar m_r\Omega^3} \sim 1.
\label{condition}
\end{equation}
Here $1/m_r = 1/m_c+1/m_v$ is the reduced mass of the effective
valence and conduction band masses $m_v$ and $m_c$ respectively.
Then the field induced effects extend a few $\hbar\Omega$ around gap and
may be viewed as being due to a combination of quantum mechanical tunneling
and multi photon processes.
Experimentally it is exceedingly difficult to fulfill the condition
(\ref{condition}) such that the effect extends over a range which can be
resolved experimentally.
The advent of the free electron laser as a source of intense coherent
radiation in the THz regime \cite{FELS} has made it possible to
enter the regime $\gamma \sim 1$,
which now has come to be known as the dynamical Franz-Keldysh (DFK) regime.
Works reporting on optical experiments in the DFK regime include
Refs.\ [\onlinecite{Kent1,cerne97,junprl,NOR98,philips}].
This has also lead to renewed theoretical interest of THz electro
optics, see e.g. Refs.\ [\onlinecite{MEIER95,JAU96,JONSSON97,CITRIN97,XU98}].
It has been shown that the physical signature of the DFK effect is enhanced
with reduced dimensionality \cite{JAU96}, which has lead to experimental
investigation of the two dimensional analogue of Yacoby's prediction in
quantum wells at low temperatures \cite{Kent1}. Of particular relevance
to the present work is the observation of optical sidebands in the
transmission spectrum at frequencies $\omega_p \pm 2n\Omega$, where
$\omega_p$ is the probe frequency and $n$ is an integer. Thus the sidebands
only appear at even multiples of the driving frequency $\Omega$ in quantum
wells. This observation can be understood as being a direct consequence of
the underlying inversion symmetry of the system \cite{JOH97}, e.g.\ the dispersion
has the property $\epsilon_n(\vec k) = \epsilon_n(- \vec k)$, $n$ is the band index. We
remark from the onset that it is assumed that the THz 
field does not induce interband coupling\cite{JOH98}.
Absence of odd sidebands has also
been observed in the presence of strong quantizing magnetic fields
\cite{junprl}.
In Ref.\ [\onlinecite{philips}] it has now been experimentally
demonstrated that odd sidebands
appear if the inversion symmetry is broken by driving the system
such that the field oscillates
in the growth direction of an asymmetric quantum well.
These observations support the assumption that the THz field only induces
intraband dynamics.
In the light of these observations and the theoretical understanding thereof
it therefore comes as a curious surprise that odd sideband formation
is observed of resonance, e.g. away from the gap,
in apparently inversion symmetric bulk samples \cite{CHIN}.
%
%
In the present work we theoretically demonstrate that inversion 
symmetry breaking results from spin-dynamic field coupling
leading to the formation of odd sidebands.
We refer to the well known relativistic effect
that a charged particle traveling with velocity $\vec v$
in an electric field $\vec E$ experiences an effective magnetic
field proportional to $\vec v\times\vec E$ which couples to the particles
spin degree of freedom via a Zeeman term, see e.g.\ Ref.\ [\onlinecite{soeffect}].
In the present case the electric field is the strong THz field driving the system.

The article is organized as follows; in Sec.\ \ref{sec:mod} we define the
theoretical model and calculate the sideband intensities neglecting the
spin-THz field coupling and show how only even sidebands appear, 
in Sec. \ref{sec:mod} we calculate effect of the spin-THz field coupling and
finally in Sec.\ \ref{sec:dis} we discuss the physical consequences of our results
and present numerical calculations for sideband generation in GaAs and InAs.
\section{THE MODEL}
\label{sec:mod}
In this section we define the model which we shall study. Our approach
is to apply non-equilibrium Green function techniques to include
the THz field nonperturbatively. This can be done analytically and
the resulting Green functions form the starting point for our subsequent
calculations, which is to account for the relativistic term
perturbatively in the Born approximation. The resulting Green functions
are then applied to determine the interband susceptibility which
describes the linear response of the system to a weak optical near
infrared probe field.

We consider a direct gap undoped bulk semiconductor subject to an
intense linearly polarized THz field, $E_{THz}(t)$. We shall assume that
effects due to the finite wave-vector are negligible, that is we assume
that the THz field is uniform. Thus the system considered remains
translationally invariant and the wave vector is still a good quantum number.
We shall model the system with a
four band model which is a generalization of the model introduced in
Ref.\ [\onlinecite{JOH98}]. A conduction band and a valence band, each carrying one
of two spin states. We start our analysis with the second
quantized Hamiltonian
\begin{equation}
H^0 = \sum_{\alpha=\{c,v\},\sigma,\vec k}
\epsilon_{\alpha\sigma}[\hbar \vec k +e\vec A(t)]
c^\dagger_{\alpha\sigma \vec k}
c_{\alpha\sigma \vec k}.
\end{equation}
Here $\sigma$ is the spin state index, $\alpha$ the band index and
$k$ is the wave vector of the state. The THz field is
introduced nonperturbatively via the vector potential
$\vec A = -\vec E\sin (\Omega t)/\Omega$. We shall assume that the THz field is
oriented in the $\hat z$-direction.
The probe has wave vector $\vec q$ and is oriented such that it forms
an angle $\theta$ to the
$\hat z$-direction, see Fig.\ \ref{figIntro}.
For simplicity, we shall assume that
the bands are parabolic, even away from the gap. We assume that
\begin{equation}
\epsilon_{c\sigma}(\vec p) = \frac{p^2}{2 m_c} + \epsilon_g
\end{equation}
for the conduction band. Here $m_c$ is effective mass for the conduction
band and $\epsilon_g$ is energy gap of the semiconductor. For the
valence band we describe the dispersion as
\begin{equation}
\epsilon_{v\sigma}(\vec p) = -\frac{p^2}{2 m_v},
\end{equation}
where the $m_v$ is the effective mass of valence band electrons.
Note that $\epsilon_{\alpha\sigma}(\vec p) = \epsilon_{\alpha\sigma}(-\vec p)$,
so the initial Hamiltonian is inversion symmetric.
Furthermore, we shall assume that the interband dipole matrix elements
$d_{cv\sigma_c\sigma_v}$ are locally independent of the wave vector,
but that they depend on the general underlying symmetries of the Brillouin
zone being probed. For instance, near a point of high symmetry like the
$\Gamma$-point we take
$d_{cv\sigma_c\sigma_v} \propto \delta_{\sigma_c\sigma_v}$, reflecting
the near gap selection rules. Away from the $\Gamma$-point these selection
rules are not in force, and we shall assume that all the matrix elements
are finite but constant. Below we shall thus consider two distinct regimes,
the {\em near gap regime} characterized by the selection rules
and the {\em far regime}, where the selection rules are not in force.

The relevant quantity to study
in order to glean the optical properties of the system, is the
interband susceptibility. The non-equilibrium causal interband
susceptibility on the Keldysh contour is given by\cite{JOH97}
\begin{equation}
\chi_{cv\beta}\vec q,t,t')
=
-\frac{i}{\hbar}
\int\frac{dk}{(2\pi)^3}
g_{c\sigma_c\sigma_c'}(\vec k+\vec q,t,t')
g_{v\sigma_v'\sigma_v}(\vec k,t',t),
\end{equation}
where $\beta = (\sigma_c\sigma_v\sigma_c'\sigma_v')$ and 
the contour ordered Keldysh Green functions are defined by
\begin{equation}
g_{\alpha\sigma\sigma'}(\vec k,t,t') = -i\langle\mathrm{T}_c[
c_{\alpha\sigma \vec k}(t)c^\dagger_{\alpha'\sigma'\vec k}(t')
]\rangle.
\end{equation}
For an overview on non-equilibrium Green functions see
\cite{noneqilibriumstuff}.
The physically relevant susceptibility is the real time
retarded susceptibility. We analytically continue the causal susceptibility
using the Langreth rules\cite{LangrethRules} and find that
\begin{equation}
\chi^r_{cv\beta}(\vec q,t,t')
=
-\frac{i}{\hbar}
\int\frac{dk}{(2\pi)^3}\Big[
g^<_{c\sigma_c\sigma_c'}(\vec k+\vec q,t,t')
g^a_{v\sigma_v'\sigma_v}(\vec k,t',t)
+
g^r_{c\sigma_c\sigma_c'}(\vec k+\vec q,t,t')
g^<_{v\sigma_v'\sigma_v}(\vec k,t',t)
\Big].
\end{equation}
In the quasiparticle picture one describes the lesser function, $g^<$,
as a distribution function $f$ times the spectral function
via $g^< = ifa$.
The spectral function is given by $a = i(g^r-g^a)$
in a non-equilibrium system. In equilibrium this generalization
becomes the familiar $a_{eq} = 2\mathrm{Im}g^r$.

In an undoped semiconductor all the valence band is occupied,
hence the distribution function is one, but no states in the conduction
band are occupied and the corresponding distribution function is zero.
We thus put $g_v^< = -(g^r_v-g^a_v)$ and $g_c^< = 0$.
The interband susceptibility becomes
\begin{equation}
\chi^r_{cv\beta}(\vec q,t,t')
=  
-\frac{i}{\hbar}
\int\frac{dk}{(2\pi)^3}
g^r_{c\sigma_c\sigma_c'}(\vec k+\vec q,t,t')
g^a_{v\sigma_v'\sigma_v}(\vec k,t',t).
\label{eq:chir}
\end{equation}
Probing the system with a weak beam
$E_p(\vec q,\omega_p t) \sim E_pe^{-i\omega_p t}$, the linearly induced
interband polarization is
\begin{equation}
P_{cv}(t) = e^2E_p\sum_{\sigma_c'\sigma_v'\sigma_c\sigma_v}
d_{cv\sigma_c\sigma_v}d_{cv\sigma_c'\sigma_v'}
\int dt' \chi^r_{cv\sigma_c\sigma_v\sigma_c'\sigma_v'}(\vec q,t,t')
e^{-i\omega_p t'}.
\end{equation}
Driving the system with THz frequency $\Omega$ the induced polarization
will be of the form
\begin{equation}
P_{cv}(t) = e^2E_p \sum_n \eta_n(\omega_p)e^{-i(\omega_p+n\Omega)t},
\end{equation}
which spectrally is a comb of oscillating dipoles giving rise to
sidebands irradiating with intensity proportional to
$I_n(\omega_p) = |\eta_n(\omega_p)|^2$ and frequency $\omega_p+n\Omega$.
The absorption of the probe is proportional to
$\mathrm{Im}\eta_0(\omega_p)$.

Neglecting relativistic effects, the Green functions obeys the
Dyson equation
\begin{equation}
\big\{
i\hbar\partial_t - \epsilon_{\alpha\sigma}(\hbar \vec k + e\vec A(t))
\big\}
g_{\alpha\sigma\sigma'}^{0r/a}(\vec k,t,t') =
\hbar\delta_{\sigma\sigma'}\delta (t-t'),
\end{equation}
with the appropriate boundary conditions.
These Green functions which include the THz field to all orders form
the starting point of our subsequent calculations. Thus our unperturbed
Green functions take the THz field into account from the onset.
The equations are readily
integrated with the results
\begin{equation}
g^{0r/a}_{\alpha\sigma\sigma'}(\vec k,t,t') = \pm\hbar\delta_{\sigma\sigma'}
\theta (\pm t \mp t')
\exp\Big\{
- i\int_{t'}^t\frac{ds}{\hbar}\epsilon_{\alpha\sigma}(\hbar \vec k + e\vec A(s))
\Big\}.
\end{equation}
Using these Green functions in order to evaluate the interband susceptibility
leads to a Gaussian integral in $\vec k$-space which we evaluate with the
result
\begin{equation}
\chi^{0r}_{cv\beta}(\vec q,\tau,T) =
\frac{e^{i\pi/4}}{(2\pi)^{3/2}}
\frac{(m_r\Omega)^{3/2}}{\sqrt{\hbar}}
e^{
- i( \hbar q^2/(2M)+\omega_f+\omega_g )\tau
}
\sum_ne^{-in\pi/2}
f_n^0(\Omega\tau)
e^{-i2n\Omega T}.
\end{equation}
Here $\tau = t-t'$, $T = (t+t')/2$, $M = m_v+m_c$, 
and we have defined
\begin{equation}
f_n^l(x) = \frac{\theta (x)}{x^{3/2}}
\sin^l(x/2)
\exp\Big(
4i\gamma \frac{\sin^2(x/2)}{x}
\Big)
J_n\Big(
\gamma \big(
\sin x - 4 \frac{\sin^2(x/2)}{x}
\big)
\Big).
\end{equation}
Other components of the susceptibility are zero.
We thus obtain that
\begin{equation}
\eta_{2n}(\omega_p) = 2
d_{cv\uparrow\uparrow}d_{cv\uparrow\uparrow}
\frac{e^{i(1-2n)\pi/4}}{(2\pi)^{3/2}}
\frac{(m_r\Omega)^{3/2}}{\sqrt{\hbar}}
\int_0^\infty
d\tau
f_n^0(\Omega\tau)
e^{i\omega_{qn}\tau},
\end{equation}
where we have used that $d_{cv\uparrow\uparrow} = d_{cv\downarrow\downarrow}$ and
we have defined $\omega_{qn} = \omega_p+n\Omega-\hbar q^2/(2M)-\omega_f-\omega_g$.
The remaining integral is readily evaluated numerically.
\subsection{PHYSICAL IMPLICATIONS}
Here several
physical results emerge. Sidebands do only appear at frequencies
$\omega_p\pm 2n\Omega$, no sidebands involving an odd number of THz photons appear.
This is a direct consequence of the inversion symmetry of the
Hamiltonian. The response is independent on the relative orientation
of the probe and the THz field, which is reflected by that
the result only depends on the magnitude of $q$.
In Fig.\ \ref{fig2} we illustrate the sideband
intensities for the first 4 sidebands as a function of the probe frequency.
These results are valid in the near zone,
and are in qualitative agreement with the experimental findings reported
in \cite{Kent1} for quantum wells. Both the probe and the irradiating
sidebands are close to real states, hence the asymmetry around $\omega_g$.
The relation $I_n(\omega) = I_{-n}(\omega+n\Omega)$ holds. The probe
or the sideband is near the main spectral feature, the band edge in the
present case. In the quantum well the main feature is the
1s exciton resonance \cite{NOR98}.

\section{THE RELATIVISTIC SPIN-THz FIELD COUPLING}
\label{sec:rel}
In this section we shall take into account
the relativistic effects. The coupling
to the spins comes physically about because a moving electron in the
presence of an electric field experiences a local magnetic field,
proportional to $\vec v\times \vec E$, which couples to its spin via a Zeeman
interaction, see e.g.\ \cite{soeffect}. This is the effect which gives rise to
spin orbit coupling in atoms and solids.
In the situation considered here the strong THz field
provides the electric field giving rise to the effect. We describe the
contribution to the Hamiltonian as
\begin{equation}
U_i(\vec k,t) = \alpha_i e\,\vec \sigma\cdot(\vec k\times \vec E(t)),
\label{eq:Ui}
\end{equation}
here $\alpha_i$, $i=\{c,v\}$, is an effective
coupling constant which depends on the material,
$\vec \sigma$ is a vector of the Pauli matrices and
$\vec E(t)$ is the intense THz field.
With $\vec E$ oriented in the $\hat z$ direction (\ref{eq:Ui}) becomes
\begin{equation}
U_i(\vec k,t) = \alpha_ie
\left(
\begin{array}{cc}
0 & k_y+ik_x \\
k_y-ik_x & 0
\end{array}\right)E_z(t).
\end{equation}
Physically the same kind of term has been studied by Raman spectroscopy
in asymmetric GaAs quantum wells\cite{jusserand}.
Then the external field is static and comes about
due to the asymmetry of the quantum well confining potential. The term breaks
time reversal symmetry and thus influences weak localization leading
to weak anti-localization \cite{pikus,hassenkam}. The important thing
for the present study is that $\alpha_c$ has been determined both
experimentally \cite{jusserand}
and theoretically \cite{pikus} for GaAs leading to
$\alpha_c \approx 5$\AA$^2$, and for InAs $\alpha_c \approx 110$\AA$^2$.
From the point of view of an effective mass picture we estimate the
coefficient for the valence band by $\alpha_v = \alpha_c(m_c/m_v)^2$.
These parameters are such that they lead to small corrections compared to the
remainder of the Hamiltonian.
We thus calculate the correction
to the Green functions to lowest order in the Born approximation,
with the result
\begin{eqnarray}
g^{r/a}_i(\vec k,t,t') &=& g^{0r/a}_i(\vec k,t,t') +
\int_{-\infty}^\infty \frac{ds}{\hbar^2}\,
g^{0r/a}_i(\vec k,t,s)U_i(\vec k,s)g^{0r/a}_i(\vec k,s,t') \\
&=&
\left(
\begin{array}{cc}
1 & \pm l_i(t,t')(k_y+ik_x) \\
\pm l_i(t,t')(k_y-ik_x) & 1
\end{array}\right)g^{0r/a}_i(\vec k,t,t'),
\end{eqnarray}
where we have introduced the quantity
\begin{eqnarray}
l_i(t,t') 
&=& 2\frac{\alpha_ieE_z}{\hbar\Omega}\cos\big(\Omega\frac{t+t'}{2}\big)
\sin (\Omega(t-t')).
\end{eqnarray}
Note that $l_i(t,t') = -l_i(t',t)$.
Again in this case it is straightforward to perform the $\vec k$ integration
in order to obtain the interband susceptibility via Eq.\ (\ref{eq:chir}).
The integrals involved are Gaussian times polynomials in $\vec k$.
We remark that higher order contributions preserve this structure as well
and are thus readily determined.
Now the relative orientation of the probe with respect to the
driving field becomes important. We assume that $\vec q$ is perpendicular to
the $\hat x$ direction.
We find that
$\chi_{cv\beta}(\vec q,\tau,T) = \chi_{cv\beta}^{0r}(\vec q,\tau,T)$
for $\beta \in \{\uparrow\uparrow\uparrow\uparrow,
\uparrow\downarrow\uparrow\downarrow,
\downarrow\uparrow\downarrow\uparrow,
\downarrow\downarrow\downarrow\downarrow\}$,
\begin{eqnarray}
\chi_{cv\beta}(\vec q,\tau,T) &=&
\frac{1}{(2\pi)^{3/2}}
\frac{2e\alpha_vq_yE_z}{\hbar\Omega}
\frac{e^{-i\pi 5/4}}{\sqrt\hbar}
\frac{m_r^{5/2}\Omega^{3/2}}{m_c}
e^{-i(\hbar q^2/(2M)+\omega_f+\omega_g)\tau}
\nonumber \\ &&\times
\cos (\Omega T)
\sum_ne^{-in\pi/2}f_n^1(\Omega\tau)
e^{-i2n\Omega T},
\end{eqnarray}
for $\beta \in \{
\downarrow\downarrow\downarrow\uparrow,
\uparrow\downarrow\uparrow\uparrow,
\uparrow\uparrow\uparrow\downarrow,
\downarrow\uparrow\downarrow\downarrow
 \}$,
\begin{eqnarray}
\chi_{cv\beta}(\vec q,\tau,T) &=&
\frac{1}{(2\pi)^{3/2}}
\frac{2e\alpha_cq_yE_z}{\hbar\Omega}
\frac{e^{-i\pi 5/4}}{\sqrt\hbar}
\frac{m_r^{5/2}\Omega^{3/2}}{m_c}
\big(1-i\frac{m_r}{m_c}\big)
e^{-i(\hbar q^2/(2M)+\omega_f+\omega_g)\tau}
\nonumber \\ && \times
\cos (\Omega T)
\sum_ne^{-in\pi/2}f_n^1(\Omega\tau)
e^{-i2n\Omega T},
\end{eqnarray}
for $\beta \in \{
\downarrow\uparrow\uparrow\uparrow,
\downarrow\downarrow\uparrow\downarrow,
\uparrow\uparrow\downarrow\uparrow,
\uparrow\downarrow\downarrow\downarrow
\}$. Other components are of second order in $\alpha$ and are negligible.
We can now find the sideband intensities from
\begin{eqnarray}
\eta_{2n}(\omega_p) &=& 2(
d_{cv\uparrow\uparrow}d_{cv\uparrow\uparrow}+
d_{cv\uparrow\downarrow}d_{cv\uparrow\downarrow})
\frac{e^{i\pi(1-2n)/4}}{(2\pi)^{3/2}}
\frac{(m_r\Omega)^{3/2}}{\sqrt{\hbar}}
\int_0^\infty
d\tau
f_n^0(\Omega\tau)
e^{i\omega_{qn}\tau}
\label{eq:even}
\end{eqnarray}
and
\begin{eqnarray}
\eta_{2n+1}(\omega_p) &=& 4d_{cv\uparrow\uparrow}d_{cv\uparrow\downarrow}
\frac{e^{-i\pi(5+2n)/4}}{(2\pi)^{3/2}}
\frac{(m_r\Omega)^{3/2}}{\sqrt{\hbar}}
\frac{m_r}{m_c}
\frac{2e(\alpha_v+\alpha_c(1-i\frac{m_r}{m_c}))q_yE_z}{\hbar\Omega}
\int_0^\infty
d\tau
( f_n^1(\Omega\tau) - i f_{n+1}^1(\Omega\tau))
e^{i\omega_{qn}\tau}
\label{eq:odd}
\end{eqnarray}
This is the main result of the present work.
\section{RESULTS AND DISCUSSIONS}
\label{sec:dis}
When $d_{cv\uparrow\downarrow} \ll d_{cv\uparrow\uparrow}$,
as is the case near the $\Gamma$-point, near resonance,
the relativistic effects are suppressed
and we obtain the results of the previous section.
The relativistic
effects, i.e.\ contributions to the sideband intensity proportional
to $\alpha_i$, manifest themselves in the far zone, away from resonance.
The relativistic term leads to the formation of
odd sidebands. Thus the inversion
symmetry has been broken and relativistic self induced local symmetry
breaking has taken place. The inversion symmetry breaking is
local because it is only detectable with finite probe wave vector $\vec q$.
We say that the symmetry breaking is global if it leads
to odd sidebands even when $q\rightarrow 0$. An example of a global symmetry
breaking
is for instance when a linear term is present in the dispersion.
The nature of the even (\ref{eq:even}) and the odd (\ref{eq:odd})
sidebands is fundamentally different.
Let $\theta$ be the angle between the probe beam, $\hat q$, and
the direction the electric field is polarized in, $\hat z$,
see Fig.\ \ref{figIntro}.
The even sidebands
do not depend on $\theta$. They can thus be regarded as a result of
pure amplitude modulations of the fundamental probe. 
For the odd sidebands however, we note that
$I_{2n+1} \propto \sin^2\theta$. The odd sidebands therefore
vanish when the probe is aligned parallel with the electric field, but is
at a maximum when they are aligned perpendicular to each other.

Taking $d_{cv\uparrow\downarrow} = d_{cv\uparrow\uparrow}$ to be
constant, we have evaluated the sideband intensities according to
(\ref{eq:even}) and (\ref{eq:odd}) using the material parameters
for the heavy hole band and conduction band for GaAs and InAs.
We have taken $\hbar\Omega = 30$meV for the calculations.
In Fig.\ \ref{fig3} we show $I_n$, $n=1,2,3,4$, for THz
intensities corresponding to $\gamma = 0.1,0.5,1.0,2.0$ as
a function of the probe frequency for GaAs.
Near resonance $I_2$ is dominating. Moving away from the resonance
$I_1$ decays much slower than the other sidebands and eventually
becomes dominating. Notice also that $I_4 > I_3$ within the region
shown. The even sidebands are more sensitive to the THz
intensity than the odd sidebands. In Fig.\ \ref{fig4} we show the
results for InAs. Qualitatively the behavior is identical except
that for the higher THz intensities we note that $I_4<I_3<I_2<I_1$
away from resonance. Quantitatively the odd sidebands are about a
factor $100$ stronger than for GaAs. In Fig.\ \ref{fig5} we show
the same sideband intensities as a function of the THz intensity
keeping the probe frequency fixed, $\omega/\Omega = 2.5,
5.0,7.5,10.0$, for GaAs. Note how $I_2$ and $I_3$ seem to rise
in a similar manner. It is noteworthy as well how feature less
$I_1$ is, in fact $I_1$ is mostly linear $\gamma$ within the
regimes we have investigated. 
In Fig.\ \ref{fig6} we show the same results for InAs.

We have considered other possible sources of inversion
symmetry breaking. Spin-orbit splitting due to the crystal field
leads to a cubic contribution in the effective Hamiltonian
for zinc-blend structures \cite{dresselhaus},
which breaks the inversion symmetry. We find however that the contribution
to the sidebands due to this term is three orders of magnitude less than
the contribution due to (\ref{eq:Ui}) considered above for GaAs and InAs.
Another possible source is the band bending at the edge of the sample due to
the pinning of the Fermi level associated with charge accumulation 
of residual charge carriers at the surface of the sample. 
Thus the relevant inversion symmetry may be 
broken if the THz field has an electric field
component perpendicular to the surface of the 
sample. We have assumed that this is not the
case.

In summary, we have studied the nonlinear generation of
optical sidebands in strongly THz driven undoped direct gap 
semiconductors. We have found that relativistic spin-THz field coupling
leads to local breaking of inversion symmetry which results in the formation
of, otherwise suppressed,  odd sidebands in the transmitted wave
of a weak near infrared interband probe. We have shown that the even sidebands
are independent on the relative orientation of the probe and the linearly polarized 
THz field, while the odd sidebands depend strongly on the relative orientation
of the probe to the THz field. The even sidebands dominate the transmission spectrum
for probe frequencies near the gap, but moving the probe away from the gap 
odd sidebands will eventually dominate. We find that the relativistic effects
are orders of magnitude stronger in InAs than in GaAs.

\begin{acknowledgements}
We thank Dr. Antti-Pekka Jauho and Dr. Simon Pedersen for useful discussions on 
various inversion symmetry breaking effects in semiconductors, 
and in particular Dr. Junichiro Kono for discussing his experimental results prior to
publication.
\end{acknowledgements}
\end{widetext}

\begin{figure}
\begin{center}
\epsfxsize=8.0cm\epsfbox{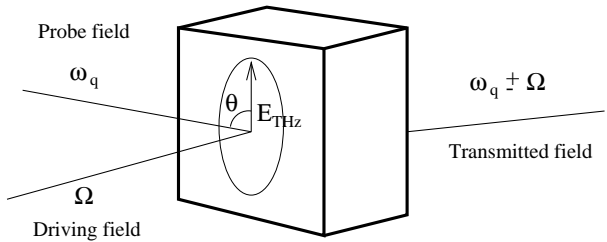}
\end{center}
\caption{
Schematic illustration of the experimental setup, showing also
the relative orientation of the probe and the linearly polarized
THz electric field.
}
\label{figIntro}
\end{figure}
\begin{figure}
\begin{center}
\epsfxsize=8.0cm\epsfbox{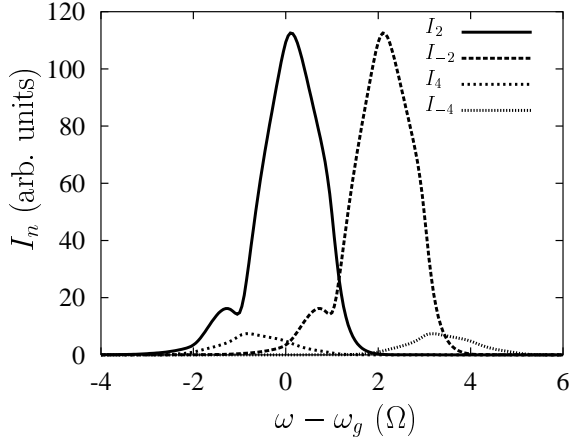}
\end{center}
\caption{
Sideband intensities $I_n$ as a function of the probe frequency
using material parameters for GaAs, $\gamma = 1$
and $\hbar\Omega = 30$meV. Only even sidebands appear due to the inversion
symmetry of the Hamiltonian. Note that
$I_n(\omega) = I_{-n}(\omega+n\Omega)$.
}
\label{fig2}
\end{figure}
\begin{figure}
\begin{center}
\epsfxsize=8.0cm\epsfbox{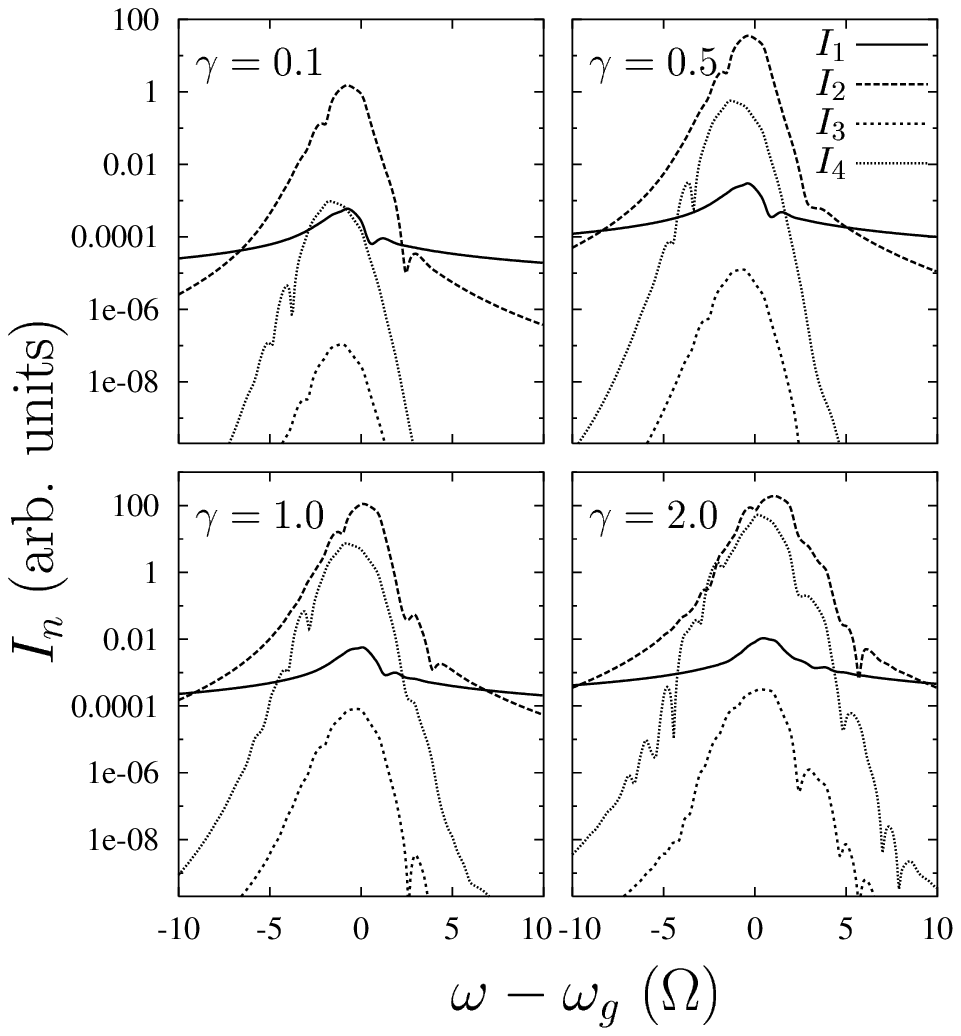}
\end{center}
\caption{
Sideband intensities $I_n$, $n\in\{1,2,3,4\}$, as a function of the probe frequency
using material parameters for GaAs including the relativistic effects. Values
of $\gamma$ are $0.1$, $0.5$, $1.0$ and $2.0$
with $\hbar\Omega = 30$meV.
}
\label{fig3}
\end{figure}
\begin{figure}
\begin{center}
\epsfxsize=8.0cm\epsfbox{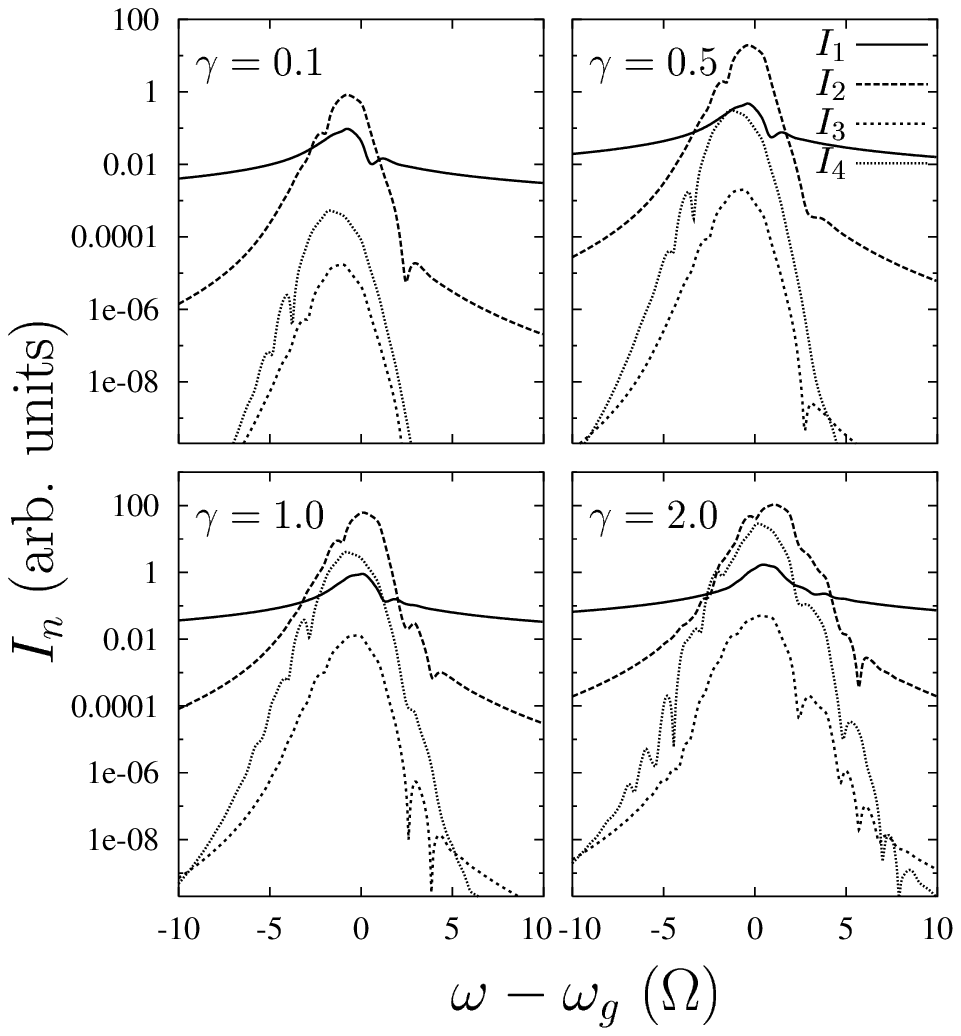}
\end{center}
\caption{
Sideband intensities $I_n$, $n\in\{1,2,3,4\}$, as a function of the probe frequency
using material parameters for InAs including the relativistic effects. Values
of $\gamma$ are $0.1$, $0.5$, $1.0$ and $2.0$
with $\hbar\Omega = 30$meV.
}
\label{fig4}
\end{figure}
\begin{figure}
\begin{center}
\epsfxsize=8.0cm\epsfbox{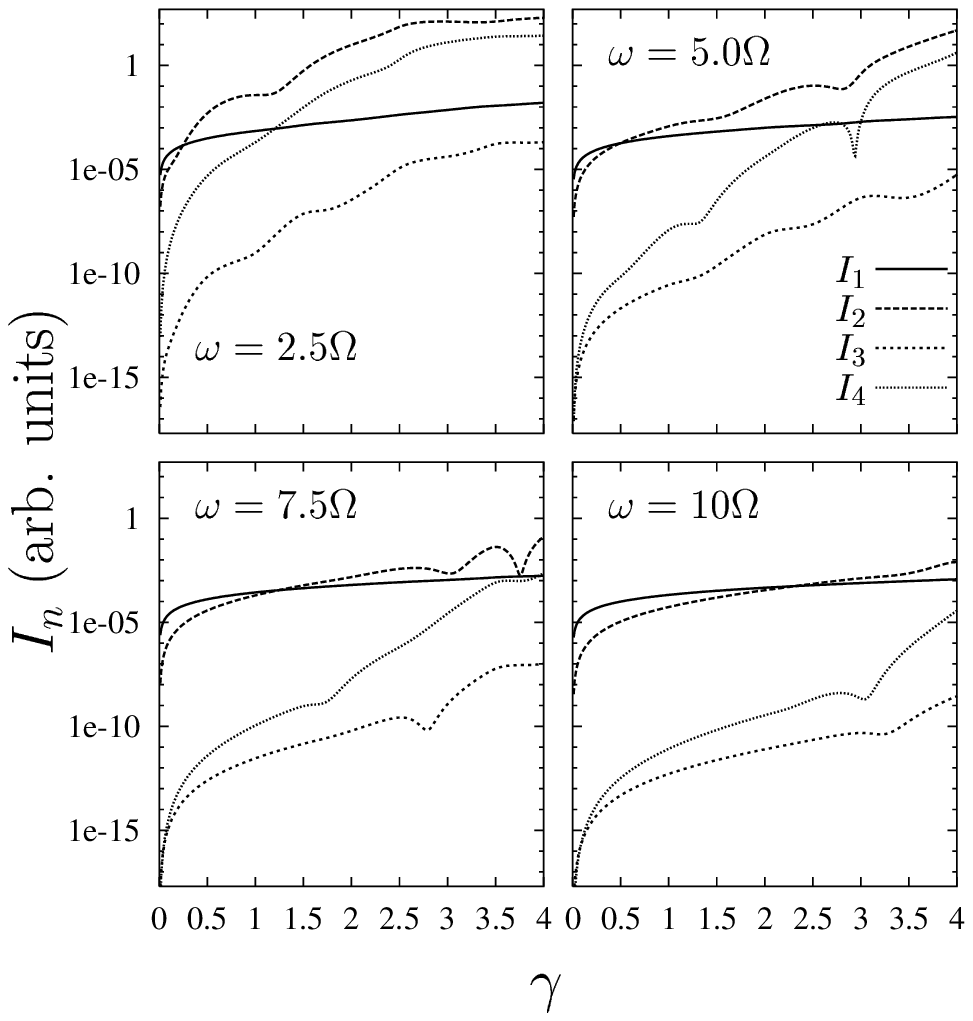}
\end{center}
\caption{
Sideband intensities $I_n$, $n\in\{1,2,3,4\}$, as a function of $\gamma$
using material parameters for GaAs including the relativistic effects. Values
of $\omega-\omega_g$ are $2.5\Omega$, $5.0\Omega$, $7.5\Omega$ and 
$10.0\Omega$ with $\hbar\Omega = 30$meV.
}
\label{fig5}
\end{figure}
\begin{figure}
\begin{center}
\epsfxsize=8.0cm\epsfbox{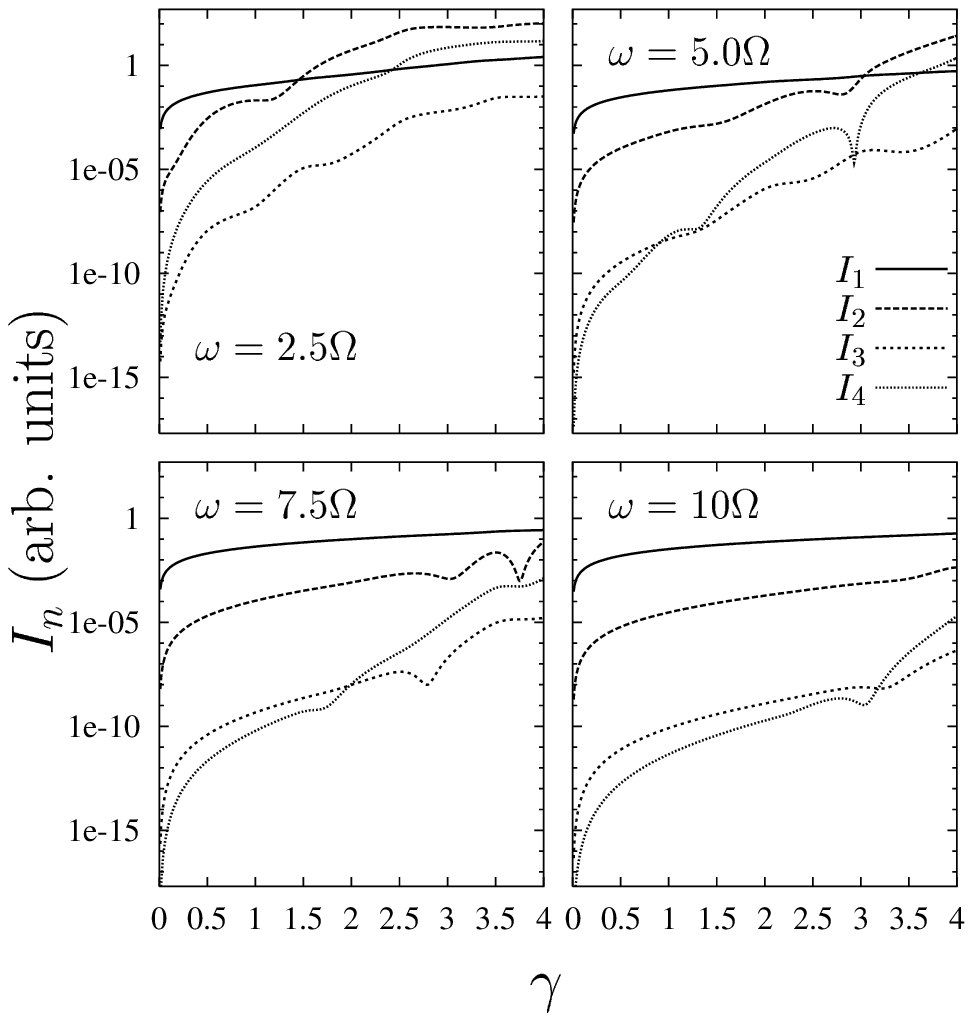}
\end{center}
\caption{
Sideband intensities $I_n$, $n\in\{1,2,3,4\}$, as a function of $\gamma$
using material parameters for InAs including the relativistic effects. Values
of $\omega-\omega_g$ are $2.5\Omega$, $5.0\Omega$, $7.5\Omega$ and 
$10.0\Omega$ with $\hbar\Omega = 30$meV.
}
\label{fig6}
\end{figure}

\begin{thebibliography}{}
\bibitem{Franz-Keldysh}
W.\ Franz, Z.~Naturforschung {\bf 13}, 484 (1958);
L.~W.\ Keldysh, Sov.~Phys.~JETP {\bf 34}, 788 (1958).
\bibitem{Yacoby}
Y.\ Yacoby, Phys.~Rev.\ {\bf 169}, 610 (1968).
\bibitem{FELS}
See, e.g., C.~A.~Brau, {\em Free-Electron Lasers} (Academic Press,
San Diego, 1990).
\bibitem{Kent1}
K.~Nordstrom {\it et~al.}, phys.~stat.~sol. (b) 204, 52 (1997).
\bibitem{cerne97}
J. Cerne {\it et~al.}, Appl. Phys. Lett. {\bf 70}, 3543 (1997).
\bibitem{junprl}  J. Kono {\it et~al.}, Phys. Rev. Lett. {\bf 79}, 1758
(1997).
\bibitem{NOR98}
K.~Nordstrom {\it et~al.}, Phys.~Rev.~Lett. {\bf 81}, 457 (1998).
\bibitem{philips}
C. Phillips {\it et~al.}, Appl. Phys. Lett. {\bf 75}, 2728 (1999).
\bibitem{MEIER95}
T.~Meier, F.~Rossi, P.~Thomas and S.~W.~Koch, Phys.~Rev.~Lett.
{\bf 75}, 2558 (1995).
\bibitem{JAU96}
A.-P.~Jauho and K.~Johnsen, Phys.~Rev.~Lett. {\bf 76},  4576  (1996).
\bibitem{JONSSON97}
L.\ J{o}nsson, M.\ M.\ Steiner, and J.\ W.\ Wilkins, Appl.~Phys.~Lett.
{\bf 70}, 1140 (1997).
\bibitem{CITRIN97}
D.\ S.\ Citrin, Appl.~Phys.~Lett. 1189 (1997).
\bibitem{XU98}
W.~Xu, J.~Phys.:Condens.~Matter {\bf 10} 10787 (1998).
\bibitem{JOH97}
K.~Johnsen and A.-P.~Jauho, phys.~stat.~sol. (a) 164, 553 (1997).
\bibitem{JOH98}
K.~Johnsen and A.-P.~Jauho, Phys.~Rev.~B {\bf 57}, 8860 (1998).
\bibitem{CHIN}
A.~H.\ Chin, J.~M. Bakker, and J. Kono, preprint (1999).
\bibitem{noneqilibriumstuff}
See, e.g., H. Haug and A.-P. Jauho,
{\em Quantum Kinetics in Transport and Optics of Semiconductors},
Springer Series in Solid-State Sciences Vol.\ 123 (Springer, Berlin, 1996).
\bibitem{LangrethRules}
D. C. Langreth, NATO Advanced Study Institute Series B, {\bf 17},
edited by J. T. Devreese and E. van Doren (Plenum, New York, 1976).
\bibitem{soeffect}
See, e.g., J. J. Sakurai,
{\em Advanced Quantum Mechanics},
(Reading, MA: Addison-Wesley, 1967).
\bibitem{jusserand} B. Jusserand {\em et~al.}, Phys. Rev. B, {\bf 51}, 4707 (1995).
\bibitem{pikus} W. Knap {\em et~al.}, Phys. Rev. B, {\bf 53}, 3912 (1996).
\bibitem{hassenkam} Tue Hassenkam {\em et~al.}, Phys. Rev. B, {\bf 55}, 9298 (1997).
\bibitem{dresselhaus} G. Dreselhaus, Phys. Rev. {\bf 100}, 580 (1955).
\end{thebibliography}
\end{document}